\documentclass[twocolumn]{jpsj3}
\usepackage{txfonts}

\title{Flux Phase as Possible Time-Reversal Symmetry Breaking 
Surface States of High-$T_C$ Cuprate Superconductors 
 }

\author{Kazuhiro Kuboki \thanks{kuboki@kobe-u.ac.jp}}
\inst{Department of Physics, Kobe University, Kobe 657-8501, Japan
} 

\abst{
At a (110) surface of a $d_{x^2-y-2}$-wave superconductor,  
superconducting order is strongly suppressed. 
In such a situation, ordered states that are forbidden in the bulk may arise.
This problem is studied for high-$T_C$ cuprate superconductors 
by treating the $t-J$ model with extended 
transfer integrals using the Bogoliubov de Gennes method. 
It is found that a flux phase with staggered currents along the surface, 
or an antiferromagnetic state can occur near the surface. 
Stability of the emergent surface states is different from system 
to system depending on the shapes of their Fermi surfaces. 
Possible relation to the experiments on the  Kerr effect that suggest
time-reversal symmetry breaking is discussed. 
 }

\begin{document}
\maketitle

\section{Introduction}

In high-$T_C$ cuprate superconductors, the competition and coexistence 
of several kinds of ordered states are important issues to clarify 
the mechanism of superconductivity.  
In high-$T_C$ cuprates, superconducting (SC) and antiferromagnetic 
(AF) states can 
occur depending on the doping rate ($\delta$). 
Previously these two states were thought to be exclusive, but 
it has recently been found that, in multilayer cuprate systems 
(in this paper the term "multilayer" will refer to three or more layers in
a unit cell), they can coexist uniformly in the same CuO$_2$ plane.\cite{Mukuda} 

Whether ordered states other than the SC and AF states exist in high-$T_C$  
cuprates is a subtle question concerning the pseudogap phase in the 
underdoped region. \cite{pgap1,pgap2}
In principle, a state that has a free energy higher than other states 
cannot occur, but it may arise if the stable ordered state 
is suppressed due to some reason.  
For example, near a (110) surface  of a $d_{x^2-y^2}$-wave  superconductor,  
SC order is strongly suppressed.
In such a case other states forbidden in the bulk, {\it e.g.} 
a flux phase may arise. 
The flux phase is a mean-field (MF) solution to the $t-J$ 
model on a square lattice that describes the low-energy electronic states of 
high-$T_C$ cuprates.\cite{Affleck,Ogata}
In this state the staggered circulating currents flow and a flux $\phi$ 
penetrates the plaquette in a square lattice.\cite{Affleck} 
Near (away from) half filling $\phi=\pm \pi$ 
($\phi \not= \pm \pi$) and the state is called the $\pi$-flux 
(staggered-flux) phase.  
(The $d$-density wave states, which have been introduced in a different context,  
have similar properties.\cite{Chakra})

Although the flux phase is only metastable except very near 
half-filling,\cite{Zhang,Hamada} it is energetically close to the SC state. 
Bejas {\it et al.} treated the $t-J$ model with 
a second-neighbor hopping term using $1/N$ expansion in the 
leading order. In this treatment, the SC and AF states are excluded, and 
they found that the flux phase is the leading instability even at high doping  
rates.\cite{Bejas} 
A mean-field (MF) calculation based on the slave-boson (SB) 
scheme\cite{Zou,Lee} 
for the $t-J$ model with second- and third- nearest-neighbor 
hopping terms (extended $t-J$ model) has been carried out  
to estimate the bare transition temperature of the flux phase, $T_{FL}$,  
assuming the absence of SC order.\cite{KKflux}
 It turned out that $T_{FL}$ may be finite even for a large doping rate 
($\delta \lesssim 0.15$). 

When the flux phase occurs near the (110) surface 
of the $d_{x^2-y^2}$-wave superconductor,  
the circulating current in the flux phase becomes a staggered current flowing 
along the surface with an amplitude decaying toward the bulk. 
This means that the time-reversal symmetry (${\cal T}$) is broken
locally near the surface. 
Experimentally, 
nonzero Kerr rotations have been observed in high-$T_C$ 
cupates,\cite{TRSB1,TRSB2,TRSB3} and it may be considered as 
the sign of the ${\cal T}$ violation.
To explain these experiments, several theories have been 
proposed.\cite{Tewari,Hosur,Oren,Varma,Persh}
We will examine whether the Kerr effect experiments 
can be explained  by the flux phase near the surface. 

Another possible surface state at the (110) surface of the 
$d_{x^2-y^2}$-wave superconductor is the AF state.  
Relative stability of the flux phase and the AF state as the emergent 
surface state depends on the dimensionality of the system as well as 
the shapes of their Fermi surfaces (FSs). 
In purely two-dimensional systems, the AF state cannot occur,  
because rotational symmetry in spin space 
would be broken spontaneously in the AF state. 
(In contrast, only discrete symmetry is broken in the flux phase.) 
The AF state can be stabilized by a weak three dimensionality that 
is always present in real systems. 
In single or double layer cuprates, three dimensionality is so weak 
that the AF state appears only near half-filling. 
For a La$_{2-x}$Sr$_x$CuO$_4$ (LSCO) system (single layer), 
the critical doping rate of the AF state  
is $\delta_c^{AF} \sim 0.02$,\cite{LSCO} and for 
a YBa$_2$Cu$_3$O$_{6+x}$ (YBCO) system 
(bilayer) $\delta_c^{AF} \sim 0.05$.\cite{YBCO}
On the other hand, in multilayer systems, the AF order survives up 
to a rather large doping region ($\delta_c^{AF} \sim 0.1$)\cite{Mukuda} 
due to the relatively strong three dimensionality. 
This implies that, in single and bilayer cuprates, the flux phase 
may be favorable as the surface state.
The surface AF order may be expected in multilayer cuprate systems 
for a doping range where only the SC order exists in the bulk. 
As we will see in the following, the shape of the FS is also responsible 
for the stability. 

In this paper we study the states near  (110) surfaces of
$d_{x^2-y^2}$-wave superconductors that are 
described by the extended $t-J$ model.
The spatial variations of the order parameters (OPs) are treated using the 
Bogoliubov de Gennes (BdG) method\cite{dG} 
based on the SBMA approximation. 
The long-range hopping terms are introduced to represent the different 
shapes of FSs for various high-$T_C$ cuprate superconductors. 
We will show that the flux phase or the AF state can occur as 
surface states, and their relative stability will be discussed. 

This paper is organized as follows. 
In Sect. 2  the model is presented and the BdG equations are derived. 
Results of numerical calculations for the surface states 
are described  in Sect. 3.
In Sect 4 the local density of states is examined.
Section 5 is devoted to summary and discussion.

\section{Bogoliubov de Gennes Equations}

We consider the $t-J$ model on a square lattice 
whose Hamiltonian is given as 
\begin{eqnarray}
\displaystyle H = -\sum_{jl\sigma} t_{jl} 
{\tilde c}^\dagger_{j\sigma} {\tilde c}_{l\sigma}
 +J\sum_{\langle j,l \rangle} {\bf S}_j\cdot {\bf S}_l, 
\end{eqnarray}
where the transfer integrals $ t_{jl}$ are finite for the first- ($t$), 
second- ($t'$), and third-nearest-neighbor bonds ($t^{''}$), 
or zero  otherwise.  
$J (>0)$ is the antiferromagnetic superexchange interaction 
and $\langle j,l \rangle$ denotes the nearest-neighbor bonds. 
${\tilde c}_{j\sigma}$ is the electron operator in Fock space without 
double occupancy, and we treat this condition using the SB 
method\cite{Zou,Lee}   
by writing ${\tilde c}_{j\sigma}=b_j^\dagger f_{j\sigma}$ under 
the local constraint $\sum_{\sigma}f_{j\,\sigma}^{\dagger}f_{j\,\sigma} 
+ b_j^{\dagger}b_j = 1$ 
at every $j$ site. Here $f_{j\sigma}$ ($b_j$) is a fermion (boson) operator  
that carries spin $\sigma$ (charge $e$); the fermions (bosons) are frequently 
referred to as spinons (holons). 
The spin operator is expressed as 
$
 {\bf S}_j = \frac{1}{2}\sum_{\alpha,\beta}
f^\dagger_{j\alpha} {\bf \sigma}_{\alpha\beta}f_{j\beta}$. 

We decouple the Hamiltonian Eq. (1) in the following 
manner.\cite{Kotliar,Suzumura,Inaba,Yamase,Yamase2} 
The bond order parameters 
$ \langle f^\dagger_{j\sigma}f_{l\sigma} \rangle$ and 
$ \langle b^\dagger_j b_l \rangle$ are introduced, and we denote 
$\chi_{jl\sigma} \equiv   \langle f^\dagger_{j\sigma}f_{l\sigma} \rangle$
for nearest-neighbor bonds. 
Although the bosons are not condensed in purely two-dimensional systems  
at finite temperature ($T$), they are almost condensed at low $T$ 
and for finite carrier doping ($\delta \gtrsim 0.05$). 
Since we are interested in the region $\delta \gtrsim 0.05$  
and low temperatures, we will treat holons as Bose condensed.
 Hence we approximate $ \langle b_j\rangle \sim \sqrt{\delta}$ and 
 $ \langle b^\dagger_j b_l \rangle \sim \delta$, and replace 
the local constraint with a global one, 
$\frac{1}{N}\sum_{j,\sigma} \langle f^\dagger_{j\sigma} f_{j\sigma}\rangle 
= 1-\delta$,  where $N$ is the total number of lattice sites. 
The spin-singlet resonating-valence-bond (RVB) OP  on the bond $\langle j,l\rangle$ 
 is given as 
$\Delta_{j,l} = \langle f_{j\uparrow}f_{l\downarrow} 
-f_{j\downarrow}f_{l\uparrow}\rangle/2$. 
 Under the assumption of the Bose condensation of holons, $\Delta_{j,l}$ 
 is equivalent to the SCOP. 
 (Then the onset temperature of $\Delta$ is the SC transition temperature, 
 $T_C$.) 
The magnetization is defined by 
$m_j = \langle n_{j\uparrow} - n_{j\downarrow} \rangle/2$
with $n_{j\sigma} = f^\dagger_{j\sigma}f_{j\sigma}$. 

The phase diagram in the plane of $\delta$ and $T$,
obtained within the SBMF approximation,
can describe the SC and AF states qualitatively.\cite{Inaba,Yamase,Yamase2}
In a quantitative sense, however, the region of the AF state is overestimated.  
The discrepancy is due to the MF treatment, and it could be remedied by 
the inclusion of fluctuations. However, to treat the fluctuations within 
the BdG calculation is beyond the scope of this work,
we introduce a phenomenological parameter $r$ ($0 < r \leq 1$)
to suppress the AF order:\cite{Brinck,Yamase3}
in the decoupling procedure of the $J$ term,  
$J \langle{\bf  S}_j\rangle \cdot {\bf S}_l$ is replaced by 
$rJ \langle{\bf  S}_j\rangle \cdot {\bf S}_l $. 

When the SC and AF order coexists, the so-called $\pi$-triplet pairing can 
occur.\cite{Fenton,Mura1,Mura2,Kyung,Apens} 
In this paper we neglect them for simplicity, because their amplitude is 
much smaller than that of the singlet SCOP. 
\begin{figure}[htb]
\begin{center}
\includegraphics[width=5.0cm,clip]{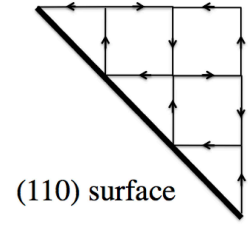}
\caption{Schematic of a (110) surface. Arrows indicate directions of currents. 
}
\end{center}
\end{figure}

We treat a system with a (110) surface (Fig.1), and 
denote the direction perpendicular (parallel) to the (110) surface as $x$ ($y$). 
The $x$ coordinate is given as $x= j_xa$  where 
$a=a'/\sqrt{2}$ with $a'$ being the lattice constant of the square lattice.
In order to describe the Flux phase and the $d$-wave SC state, 
$\chi^{(\pm)}_{j\sigma} \equiv \chi_{j,j+x\pm y,\sigma}$ and 
$\Delta^{(\pm)}_j \equiv \Delta_{j.j+x\pm y}$ are defined.
We assume that the system is uniform along the $y$ direction, 
and consider the spatial variations of OPs only in the $x$ direction. 
By imposing the periodic boundary condition for the $y$ direction, 
the Fourier transformation for the $y$ coordinate is 
performed.\cite{KKBdG,Tanuma,Zhu,KKBdG2}
(Hereafter we write $j_x$ simply as $j$.)
Then the mean-field Hamiltonian is written as follows 
\begin{equation}
\displaystyle 
{\cal H}_{ MF} = \sum_k\sum_{j,l}\Psi_j^\dagger(k) 
{\hat h}_{jl}(k) \Psi_l(k), 
\end{equation} 
with $\displaystyle \Psi_j^\dagger(k) 
= \big(f_{j\uparrow}^\dagger(k), f_{j\downarrow}(-k)\big)$, and 
$k$ is the wave number along the $y$ direction. 
The matrix ${\hat h}_{ij}(k)$ is given as
\begin{equation}
\displaystyle {\hat h}_{jl}(k) = 
\left (\begin{array}{cc}
\xi_{jl\uparrow}(k)  & F_{jl}(k)  \\
F_{lj}^{*}(k) & - \xi_{jl\downarrow}(k)
\end{array}\right ), 
\end{equation}
where
\begin{equation}\begin{array}{rl}
\xi_{jl\uparrow}(k) = & \displaystyle 
-\delta_{j,l}\big[\mu - rJ(m_{j-1}+m_{j+1})
+2t'\delta \cos 2ka \big] \\ 
& \displaystyle -\delta_{j,l-1} 
\big[2t\delta \cos ka  + \frac{J}{2}[
 (\chi^{(+)}_{j\downarrow}+\frac{1}{2}\chi^{(+)}_{j\uparrow})e^{ik} \\
 & \displaystyle
 +  (\chi^{(-)}_{j\downarrow}+\frac{1}{2}\chi^{(-)}_{j\uparrow})e^{-ik}]
 \big] \\
& \displaystyle  -\delta_{j,l+1}
\big[2t\delta \cos ka + \frac{J}{2}[
 (\chi^{(+)}_{l\downarrow}
 +\frac{1}{2}\chi^{(+)}_{l\uparrow})^*e^{-ik} \\
 & \displaystyle
 + (\chi^{(-)}_{l\downarrow}
 +\frac{1}{2}\chi^{(-)}_{l\uparrow})^*e^{ik}]
 \big] \\
& \displaystyle -(\delta_{j,l+2}+\delta_{j,l-2})(t'\delta + 2t''\delta \cos 2ka),  
\\
& \\
\xi_{jl\downarrow}(k) = & \displaystyle  
-\delta_{j,l}\big[\mu + rJ(m_{j-1}+m_{j+1})
+2t'\delta \cos 2ka \big] \\ 
& \displaystyle -\delta_{j,l-1} 
\big[2t\delta \cos ka  + \frac{J}{2}[
 (\chi^{(+)}_{j\uparrow}+\frac{1}{2}\chi^{(+)}_{j\downarrow})^*e^{ik} \\
 & \displaystyle
 +  (\chi^{(-)}_{j\uparrow}+\frac{1}{2}\chi^{(-)}_{j\downarrow})^*e^{-ik}]
 \big] \\
& \displaystyle  -\delta_{j,l+1}
\big[2t\delta \cos ka + \frac{J}{2}[
 (\chi^{(+)}_{l\uparrow}
 +\frac{1}{2}\chi^{(+)}_{l\downarrow})e^{-ik} \\
 & \displaystyle
 + (\chi^{(-)}_{l\uparrow}
 +\frac{1}{2}\chi^{(-)}_{l\downarrow})e^{ik}]
 \big] \\

& \displaystyle -(\delta_{j,l+2}+\delta_{j,l-2})(t'\delta + 2t''\delta \cos 2ka),  
\\
& \\
F_{jl}(k) = & \displaystyle  \frac{3J}{4} 
\big[\delta_{j,l-1}(\Delta^{(+)}_j e^{ika} + \Delta^{(-)}_j e^{-ika}) \\
+ & \displaystyle \delta_{j,l+1} (\Delta^{(+)}_l e^{-ika} + \Delta^{(-)}_l e^{ika})\big],    
\end{array}\end{equation}
with $\mu$ being the chemical potential, 

We diagonalize the mean-field Hamiltonian by solving the
following BdG equation for each $k$, 
\begin{equation} 
\sum_l {\hat h}_{jl}(k) 
\left (\begin{array}{cc}
u_{ln}(k)   \\
v_{ln}(k)
\end{array}\right )
= E_n(k) 
\left (\begin{array}{cc}
u_{jn}(k)   \\
v_{jn}(k)
\end{array}\right ),   
\end{equation} 
where $E_n(k)$ and ($u_{jn}(k)$, $v_{jn}(k)$) are the energy eigenvalue
and the corresponding eigenfunction, respectively, for each
$k$. The unitary transformation using ($u_{jn}(k)$, $v_{jn}(k)$) diagonalizes
the Hamiltonian ${\cal H}_{MF}$, and the OPs and the spinon number 
at the $j$ site can be written as, 
\begin{equation}\begin{array}{rl}
\langle n_{j\uparrow}\rangle = & \displaystyle 
\frac{1}{N_y} \sum_k \sum_{n=1}^{2N_x}
\big| u_{j,n}(k) \big|^2 f(E_n(k)), 
\\
\langle n_{j\downarrow}\rangle = & \displaystyle
\frac{1}{N_y} \sum_k \sum_{n=1}^{2N_x} 
\big| v_{j,n}(k) \big|^2 \big[1-f(E_n(k))\big],  
\\
\chi^{(\pm)}_{j\uparrow} = 
& \displaystyle \frac{1}{N_y} \sum_k \sum_{n=1}^{2N_x}
e^{\mp ika} u^*_{j+1.n}(k)u_{j,n}(k) f(E_n(k)), 
\\
\chi^{(\pm)}_{j\downarrow} = 
& \displaystyle \frac{1}{N_y} \sum_k \sum_{n=1}^{2N_x}
e^{\pm ika} v_{j+1.n}(k)v^*_{j,n}(k) \big[1-f(E_n(k))\big], 
\\
\Delta^{(\pm)}_j = & \displaystyle \frac{1}{4N_y} \sum_k \sum_{n=1}^{2N_x}
\big[e^{\mp ika}  u_{j,n}(k) v^*_{j+1,n}(k) 
\\ & \displaystyle 
+ e^{\pm ika} u_{j+1,n}(k) v^*_{j,n}(k)\big]
\tanh\Big(\frac{E_n(k)}{2T}\Big), 
\end{array}\end{equation}
where $N_x$ ($N_y$) and $f$ are the number of lattice sites along 
the $x$ ($y$) direction and the Fermi distribution function, respectively. 
The $d$- and $s$-wave SCOPs are obtained by combining $\Delta^{(\pm)}$s: 
$\Delta_d(j) = (\Delta^{(+)}_j - \Delta^{(-)}_j + \Delta^{(+)}_{j-1} 
- \Delta^{(-)}_{j-1})/4$ and 
$\Delta_d(j) = (\Delta^{(+)}_j + \Delta^{(-)}_j + \Delta^{(+)}_{j-1} 
+ \Delta^{(-)}_{j-1})/4$.

\section{Surface States}

In this section we present the results of numerical calculations for surface states. 
Spatial variations of the OPs near (110) surfaces are determined 
by solving the BdG equations, and we restrict ourselves to the case of $T < T_C$,   
namely, we do not consider the pseudogap phase. 
In numerical calculations, we diagonalize the Hamitotonian 
${\cal H}_{MF}$ with the OPs substituted in matrix elements, 
and the resulting eigenvalues and eigenfunctions 
are used to recalculate the OPs. This procedure is iterated until the 
convergence is reached. 
For the system size, $N_x=200$ and $N_y=100$ are used throughout. 

First we study the LSCO system. 
The transfer integrals are chosen to reproduce the FS of the 
LSCO system, $t/J=4$, $t'/t=-1/6$, and $t''=0$,\cite{tana}  and 
$\delta=0.10$ and $T=0.01J$ are used. 
In the LSCO system,  the region of the AF state is very narrow  
($\delta \lesssim 0.02$), and we do not consider it. 
In Fig.2, the spatial variations  of the OPs are shown. 
It is seen that  the $d$-wave SCOP $\Delta_d$ is suppressed near the surface 
($x=0$),  and the imaginary part of the bond OP, Im$\chi$, 
is finite in this region. 
(In the absence of magnetic order, $\chi_\uparrow = \chi_\downarrow$.)
This mean that the flux phase arises as a surface state and the time-reversal 
symmetry is broken locally near the surface. 
The critical value of the doping rate for the appearance of the flux phase, 
$\delta_c$, in the LSCO system is $\delta_c \sim 0.20$. 
In the SBMA calculation for a uniform system, 
the bare transition temperature of the flux phase, $T_{FL}$,  
vanishes at $\delta \sim 0.15$.\cite{KKflux}
Thus, $\delta_c$ in the BdG calculation is larger than that for the 
uniform system, because the incommensurate flux order that is not 
taken into account in the latter may be possible. 

Next we examine the YBCO system. 
We use a simplified parametrization of the transfer integrals 
neglecting  bilayer splitting of the FS, 
$t/J=4$, $t'/t=-1/6$, and $t''/t=1/5$,\cite{tana} and 
$\delta=0.06$ and  $T=0.01J$ are used.  
(We do not consider the AF state as in the LSCO system.)
In Fig.3 the spatial variations of the OPs are shown, and we see that the flux phase 
and hence the ${\cal T}$ violation also occur in this system. 
The value of Im$\chi$ that characterizes the flux phase is 
smaller compared with that in Fig.2, though the doping rate is smaller here. 
For the YBCO system, $\delta_c \sim 0.10$ in the BdG calculation, while 
$\delta_c \sim 0.08$ in the MF calculation for the uniform system. 

\begin{figure}[htb]
\begin{center}
\includegraphics[width=6.5cm,clip]{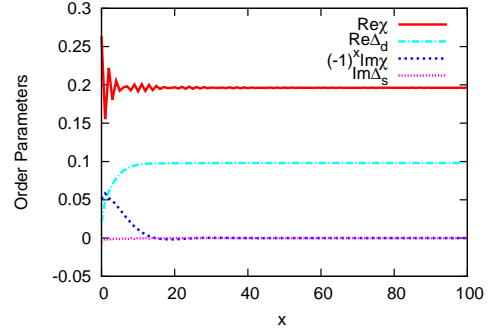}
\caption{(Color online) Spatial variations of the OPs for the LSCO system 
($t/J=4$, $t'/t=-1/6$, $t''=0$, $\delta=0.10$, and $T=0.01J$). 
Here $x$ is measured in units of lattice spacing $a$. 
Note that all OPs are nondimensional.
}
\end{center}
\end{figure}
\begin{figure}[htb]
\begin{center}
\includegraphics[width=6.5cm,clip]{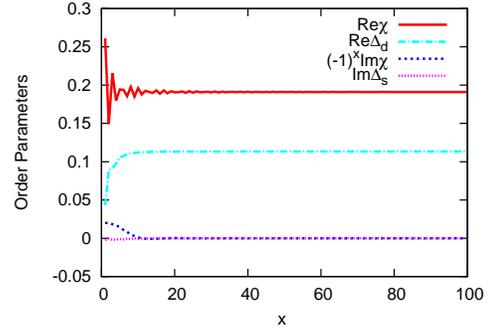}
\caption{(Color online) Spatial variations of the OPs for the YBCO system 
 ($t/J=4$, $t'/t=-1/6$, $t''/t=1/5$, $\delta=0.06$, and $T=0.01J$).
}
\end{center}
\end{figure}

The Flux phase arises in a rather large doping range for  the LSCO 
compared to the YBCO system.  
The reason for the difference is as follows. 
Flux phases are characterized by the imaginary part of 
the bond OP, Im$\chi$.  Self-consistency equations for the uniform system  
show that the expression of Im$\chi$ has a form factor 
$(\gamma_k^{(-)})^2$ ($\gamma_k^{(\pm)} 
\equiv \cos k_x \pm \cos k_y$).\cite{Zhang,Hamada,KKflux} 
Therefore, if $(\gamma_k^{(-)})^2$ is large near the FS,  the flux phase should be 
favored. \cite{KKflux}
The FS of the LSCO (YBCO) system is favorable (unfavorable) in this sense.

When the flux phase occurs,  it is seen that the imaginary part of the $s$-wave SCOP becomes finite near the surface, and thus the SC state has a $(d \pm is)$-wave 
symmetry. However, the absolute value of Im$\Delta_s$ is very small 
(of the order of $10^{-3}$), and it is considered to be driven by the flux phase order. 
The absence of the $s$-wave SCOP, when the flux phase is not present, 
 can be understood by using the Ginzburg-Landau (GL) theory.  
The coefficient of the quadratic term of the $d$-  ($s$-) wave SCOP, 
$\alpha_d$ ($\alpha_s$),  
in the GL theory has been derived microscopically from the $t-J$ model.\cite{KKGL} 
The $d$- ($s$-) wave SCOP is characterized by the form factor 
$(\gamma_k^{(-)})^2$ ($(\gamma_k^{(+)})^2$), and if 
$(\gamma_k^{(-)})^2$ ($(\gamma_k^{(+)})^2$) is large  
near the FS,  the $d$- ($s$-) wave SC state is favored. 
For the parameters used in the present BdG calculations,  
$\alpha_s$ is positive both for the LSCO and YBCO systems at least 
for $T  \geq 10^{-4}J$. 
In general, surface scatterings may induce Re$\Delta_s$, but not Im$\Delta_s$.
Moreover, for (110) surfaces this contribution vanishes by symmetry.\cite{SigUe}  
Therefore, even when $\Delta_d$ is suppressed near the (110) surface, 
$\Delta_s$ would not be induced, because no energy gain is expected. 

Although we have considered only (110) surfaces, we may expect 
${\cal T}$ violation for surfaces with other types of orientations.
In real systems surfaces will not be so smooth,  then there 
may be small domains where the angle of crystal axes is 
45$^\circ$ to the surface.  
In this case the flux phase would appear leading to ${\cal T}$ violation 
locally in these domains.  
For a surface perpendicular to the $c$ axis, grain boundaries 
could also be the origin of the flux phase order. 
  
The current along the surface ($y$ axis) is proportional to Im$\chi$, 
\begin{equation}
\displaystyle 
J_y(j) = \frac{\sqrt{2} \pi t \delta}{\phi_0} 
\sum_\sigma {\rm Im} \chi^{(+)}_{j\sigma}, 
\end{equation} 
with $\phi_0=h/2e$ being the flux quantum.  
(In principle, there is a term proportional to the vector potential in $J_y$, 
but we neglect it for simplicity.)
Then the staggered current $J_y$ flows in the region 
where the flux phase order is present.  
The magnetic field at the surface can be roughly estimated by
$
B_z(x=0) = \mu_0 \int_0^\infty dx J_y(x) 
= \mu_0 a\sum_j J_y(j). 
$
For the parameters used above (corresponding to those in Figs. 2 and 3),
$B_z(x=0)$ is of the order of 1-10 G. 
The estimated value of $B$ is small but finite, then it could lead to a finite 
but small Kerr angle observed experimentally.

The Kerr angles at the opposite surfaces have 
the same sign,\cite{TRSB3} in contradiction to uniform ${\cal T}$ violation. 
In the present theory, 
since the ${\cal T}$ violation occurs only near the surface, 
the signs of the Kerr angles at the opposite sides of the sample can be arbitrary. 
Therefore it may give a simple explanation for the experimental finding.

Theoretically, the surface flux order can occur only below $\delta_c \sim 0.10$ 
for the YBCO system, and this doping rate is less than the value for which 
the Kerr rotation is observed in the SC region.\cite{TRSB1} 
The reason for the discrepancy could be due to the fact that we have used 
the single-layer (single-band) $t-J$ model. 
If the bilayer $t-J$ model is employed, there are multiple FSs, and 
the condition for the occurrence of the flux phase may be changed. 

In contrast to the LSCO and YBCO systems, 
the AF state survives up to large $\delta$ in multilayer cuprate 
superconductors in which the coexistence of the AF and SC states 
has been found. 
For example, $T_N$ vanishes at $\delta_c^{AF} \sim 0.1$ 
for five-layer cuprate systems.\cite{Mukuda} 
In a state with $\delta \gtrsim \delta_c^{AF}$, 
AF order is suppressed by SC order, though the bare  
transition temperature of the former, $T_N^{bare}$,  is still finite. 
When SC order is suppressed near the (110) surface, 
there is a competition between the AF state and the flux phase. 
Here we use the single-layer $t-J$ model for simplicity 
($t/J=4$, $t'=t''=0$), and choose $r=0.66$. 
For this value of $r$,  $T_N$  vanishes 
at $\delta_c^{AF} \sim 0.10$. 
In Fig. 4, the results for $\delta=0.11$ ($> \delta_c^{AF}$)
and $T=0.01J$ are shown.
It is seen that the staggered magnetization $M (=(-1)^x m)$ 
is finite near the surface,  
while Im$\chi =0$ everywhere.  This means that the AF state is 
more robust than the flux phase in this system. 
For larger values of $\delta$,  surface AF order diminishes, 
and the flux phase may appear. 
The above results indicate that the emergent surface state may be 
different from system to system, depending on the shapes of the FSs. 

\begin{figure}[htb]
\begin{center}
\includegraphics[width=6.5cm,clip]{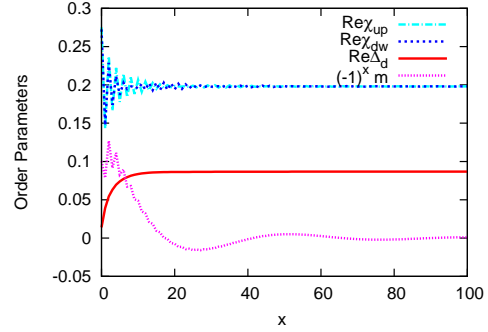}
\caption{(Color online) Spatial variations of the OPs for the system near the 
coexistence of the SC and AF states 
($t/J=4$, $t'=t''=0$, $r=0.66$, $\delta=0.11$, and $T=0.01J$). 
}
\end{center}
\end{figure}

\section{Local Density of States}

The local density of states (LDOS) at the  $j$ site
is given as 
\begin{equation}\begin{array}{rl}
N_\uparrow(j,E) = & \displaystyle \frac{1}{N_y} \sum_k \sum_{n=1}^{2N_x}
\big| u_{j,n}(k) \big|^2 \delta\big(E-E_n(k)\big), 
\\
 N_\downarrow(j,E) = & \displaystyle  \frac{1}{N_y} \sum_k \sum_{n=1}^{2N_x}
\big| v_{j,n}(k) \big|^2 \delta\big(E+E_n(k)\big), 
\end{array}\end{equation}
where $\uparrow$ and $\downarrow$ denote the spin directions.  

Figure 5 shows the LDOS for the LSCO and YBCO systems  
at the surface. 
The parameters are the same as those used in  Figs. 2 and 3.  
It is  seen that each LDOS has split peaks below and above zero energy.  
The splitting of the peaks for the LSCO system is larger than that 
for the YBCO system, reflecting the fact that Im$\chi$ is larger than that 
of the latter.

Covington {\it et al.} observed  the peak splitting of the zero bias 
conductance in $ab$-oriented YBCO/insulator/Cu junctions,
and it is considered as the sign of 
${\cal T}$ violation.\cite{Coving} 
This ${\cal T}$ violation  has been explained by 
the occurrence of an additional SCOP near the junction other than 
the bulk $d$-wave SCOP.\cite{Fogel} 
If we consider that this peak splitting is due to the surface flux phase,  
the theoretical peak-to-peak separation for YBCO in Fig. 5 ($\lesssim 1$ meV) 
is about half of those observed experimentally, and so it may be considered 
to be in qualitative agreement. 
However, the doping rate used here is lower than that of the sample 
for the tunneling experiment ($T_C=89$K),\cite{Coving} 
because ${\cal T}$ violation is limited to $\delta < \delta_c \sim 0.10$  
theoretically. 
In order to see whether surface flux phase can explain the ${\cal T}$ violation 
in YBCO/insulator/Cu junctions, 
quantitative calculations  employing the bilayer $t-J$ model will be necessary.  

The LDOS for the system with surface AF order is shown in Fig. 6. 
The parameters are the same as those used in Fig. 4. 
Here, the LDOS is different for spin directions because of magnetic order. 
This behavior may be detected by spin-dependent STM/STS experiments. 

\begin{figure}[htb]
\begin{center}
\includegraphics[width=6.5cm,clip]{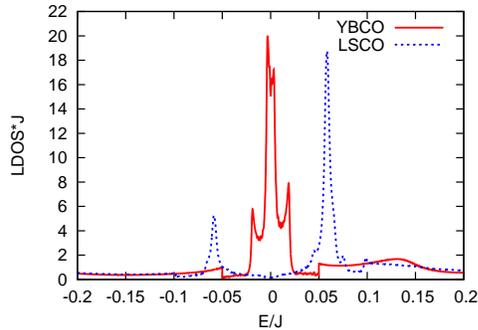}
\caption{(Color online) LDOS  at surfaces of the LSCO and YBCO systems. 
The parameters are the same as those used in Figs. 2 and 3. 
}
\end{center}
\end{figure}
\begin{figure}[htb]
\begin{center}
\includegraphics[width=6.5cm,clip]{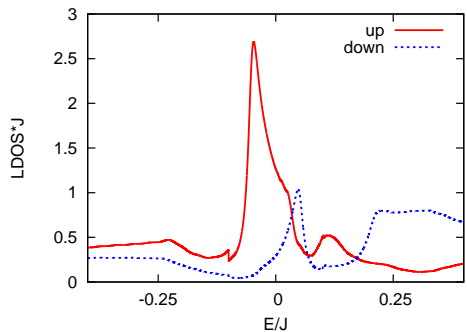}
\caption{(Color online) LDOS  at the surface of the system 
near the coexistence of the SC and AF states.  
The parameters are the same as those used in Fig. 4. 
}
\end{center}
\end{figure}

\section{Summary and Discussion}

We have studied the states near the (110) surfaces of 
$d_{x^2-y^2}$- wave superconductors that are described by 
the extended $t-J$ model. 
Since SC order is strongly suppressed there, the flux phase may 
occur and it can exist even for a large doping rate ($\delta \lesssim 0.20$), 
if the Fermi surface of the system is favorable to this state. 
Since the flux phase breaks ${\cal T}$, this may explain the results of 
experiments on the Kerr effect.\cite{TRSB1,TRSB2,TRSB3} 
In the present theory, the ${\cal T}$ violation occurs only near the surface, 
and the surface states at the opposite sides of the sample are independent. 
Then it may give a simple explanation of the experimental 
finding that the signs of the Kerr angles at the opposite 
surfaces  are the same.\cite{TRSB3}
The AF state may also appear near the (110) surface, and relative stability 
of the emergent surface states may be different from  system to system 
depending on their FSs. 

In this paper we did not consider the pseudogap phase 
({\it i.e.},  $T > T_C$ in the underdoped region). 
Experimentally, nonzero Kerr rotations are observed not only in the SC 
state but also in the pseudogap phase.\cite{TRSB1,TRSB2,TRSB3} 
In order to understand this, 
we should include the dynamics of holons,  and the effect of 
fluctuations around the SBMA solution ($U(1)$ gauge fluctuations) 
must be examined.  
Moreover, in experiments on the YBCO system, nonzero Kerr rotations 
are observed for doping rates higher than $\delta_c$ that 
we have obtained theoretically.\cite{TRSB1} 
This could be due to the fact that we have treated only 
the single-layer (single-band) $t-J$ model. 
In bilayer or multilayer models, there are multiple FSs, and 
the condition for the occurrence of the flux phase and the AF state  
may be changed quantitatively. 
Whether this scenario is correct or not can be checked by 
carrying out similar calculations 
employing the bilayer $t-J$ model. 
This problem will be studied separately in the future.


\begin{acknowledgment}
 The author thanks M. Hayashi, M. Mori, and H. Yamase for useful discussions. 
This work was supported by JSPS KAKENHI Grant Number 24540392.
\end{acknowledgment}


\end{document}